\documentclass[twocolumn,twoside,slac_two]{revtex4}
\pdfoutput=1
\usepackage[pdftex]{graphicx}
\usepackage{fancyhdr}
\def\lsim{\;\centeron{\raise.35ex\hbox{$<$}}{\lower.65ex\hbox
{$\sim$}}\;}
\def\gsim{\;\centeron{\raise.35ex\hbox{$>$}}{\lower.65ex\hbox
{$\sim$}}\;}
\def\pom{I\!\!P}
\begin{document}

\title{Accelerator Data for Cosmic Ray Physics}


\author{M.G. Albrow}
\affiliation{FNAL, Batavia, IL 60510, USA}

\begin{abstract}
I present selected examples of accelerator data, mainly from hadron colliders,
 that are relevant for understanding cosmic ray showers. I focus on the
 forward region, $x_{Feynman} > 0.05$, where high energy data are scarce, since
 the emphasis in collider physics became high-$p_T$ phenomena.
\end{abstract}

\maketitle

\thispagestyle{fancy}


\section{Introduction}
I give a brief tour of data from accelerator-based experiments on particle production that
are most relevant for understanding high energy cosmic ray showers. The database of particle production in
hadron collisions is vast, and I will have to ignore most of it. I decide to focus on the highest energy
laboratory collisions, namely hadron colliders, and the highest production cross sections, mostly forward
particles. It is therefore far from being a complete review, with a strong bias towards my taste (and
knowledge), for which I apologize to nearly everyone. Much of what I leave out is the very high transverse
momentum ($p_T$) small cross section ($\sigma$) physics that now dominates the hadron collider field (weak
vector bosons, top quarks, very high $p_T$ jets, supersymmetry and Higgs boson searches, etc.). We can
suppose these have no relevance to cosmic ray shower development, although when it comes to interactions of 
primaries with energy
$E > 10^{18}$ eV, nobody knows. I do not attempt to cover new results to be reported at this symposium by other
speakers (from experiments at RHIC; MIPP, CDF, and DZero at Fermilab; ATLAS, CMS, LHCb at the LHC); this is
an introductory talk, not a summary. My final ``cut'' is Feynman-$x$, 
$x_F = p_L/p_{beam} \gtrsim$ 0.05, after which my material is tractable! An alternative longitudinal momentum
variable is \emph{longitudinal rapidity}, or just rapidity, $y = \frac{1}{2}\mathrm{ln}\frac{(E+p_L)}{(E-p_L)}$. A Lorentz boost along the
longitudinal axis just adds a constant to all $y$-values, so rapidity differences are invariant. In a $pp$ collision the
total $y$ interval is $\Delta y = \mathrm{ln} (2 p_{beam}/m_p)$. 

Why is accelerator data important for very high energy cosmic ray physics? If the atmosphere totally contains the energy of 
a showering cosmic ray, it is a homogeneous calorimeter, and \emph{to some degree} the total fluorescence light is proportional to
the incoming energy. The fluorescence (scintillation) light is a measure of the total path length of all 
charged particles, including those in electromagnetic
showers from $\pi^0 \rightarrow \gamma\gamma \rightarrow e^+$ and $e^-$, which also give Cherenkov 
radiation. We
want to know not only the energy and direction of the primaries, but their identity (protons, iron, something else?). Measuring the
lateral and longitudinal profiles of the showers, and the muon content at some depth, are the main paths to this understanding, but
they are more sensitive to shower models. The measured shower energy will depend on the longitudinal profile (how much goes into
the ground?). Muons come mostly (but not exclusively) from $\pi^{\pm}$ and $K^{\pm}$ decays. In hybrid experiments 
such as AUGER
that can measure all these parameters for \emph{some} of the showers, does everything fit together? To answer this question it is
essential to have shower simulation models, such as \textsc{kaskade, hpdm, venus, sibyll, qgsjet, ...}. These models give what we
\emph{expect} about VHE interactions, and they should be tested against accelerator data where possible, but it is a far
extrapolation from ISR or even Tevatron energy ($E_{\equiv} = 2 \times 10^{15}$ eV) to $10^{20}$ eV! 

Let us remember the history of
energy steps in accelerator physics, from PS/AGS ($2.8\times 10^{10}$ eV) $\Longrightarrow$ 
ISR ($2.1\times 10^{12}$ eV)$\Longrightarrow$ Sp$\mathrm{\bar{p}}$S/Tevatron ($2 \times 10^{15}$ eV) 
$\Longrightarrow$ LHC ($10^{17}$ eV), with
striking new physics coming in at each step (and anticipated for the LHC). Step 1: PS $\Longrightarrow$ ISR, gave us rising $\sigma_T$,
high $p_T$ hadrons and jets, charm and beauty, and high mass diffraction. Step 2: ISR $\Longrightarrow$ Sp$\mathrm{\bar{p}}$S/TeV, gave us more 
dramatic high $p_T$ jets, W and Z bosons, prolific heavy flavor production and top quarks. 
Step 3: TeV $\Longrightarrow$ LHC, 
\emph{must} give us abundant $p_T \sim$ TeV-scale jets and top quarks, open the electroweak sector with 
abundant $W$ and $Z$ and probably Higgs bosons, and quite likely (let us hope for) 
supersymmetric particles and/or other
new particles or phenomena. Note that the famous ``knee'' in the cosmic ray spectrum is in the 
middle of the TeV $\Longrightarrow$ LHC step, and there have been suggestions~\cite{white,knee} that it is caused by a change
in the nature of the interactions (which would have to be dramatic!).
Even so, the step LHC $\Longrightarrow$ VHECR is another stretch from $10^{17} \Longrightarrow
10^{20}$ eV, with room for more surprises. 

Whether or not there are surprises, consider the extrapolations of fits to existing data on basic quantities such as the mean
charged multiplicity $\langle n_{ch} \rangle$, or mean (particle) transverse momentum $\langle p_T \rangle$, as given by the
\textsc{dpmjet II.5} generator~\cite{dpmjet}. As $\sqrt{s}$ rises from 0.1 TeV to 1000 TeV, $\langle n_{ch} \rangle$ is expected to
grow from about 12 to about 170, and $\langle p_T \rangle$ from 0.4 GeV/c to slightly over 1 GeV/c; these are large changes,
and will remain large even with constraints from LHC data.

I find it striking that most of the accelerator data passing my ``forward-looking cuts'' (e.g. $x_F > 0.05$), where most of the particles
and energy are, comes from the ISR, with some from RHIC at $\sqrt{s}$ = 500 GeV, but very little else, 
so the extrapolation is nearly eight orders of magnitude in eV. This is because the high-$p_T$
sector (probing quarks and gluons at a scale $\sim 10^{-3}$ fm, and top quarks) and the electroweak sector ($W,Z,H(?)$) dominated
the post-ISR program. To a good approximation 
the only post-ISR detectors able to measure particle spectra with $x_F > 0.05$
were small trackers in ``Roman pots'' able to detect $x_F \gtrsim$ 0.9 
diffractively scattered (anti-)protons. Largely this was due to lack of interest (although the cosmic ray community was interested), as
well as the difficulty of making a small angle spectrometer fit in a very limited space. Later I will speculate on whether such a
spectrometer could be made for the LHC.

\section{CERN Intersecting Storage Rings, ISR}
  The Intersecting Storage Rings at CERN was the first hadron collider~\cite{prisr}. The first collisions
  occured in February 1971, and I hope CERN will celebrate the 40th anniversary next year. It was a
  remarkable machine, with two independent rings crossing at eight intersection regions. The beams were
  continuous flat ribbons, not bunched as in all other colliders, proton beam currents were to reach
  60 amps, and luminosities above 10$^{32}$ cm$^{-2}$s$^{-1}$, a record that held for over 20 years! Not
  only protons, but antiprotons, deuterons and $\alpha$-particles could be stored and collided in any
  combination ($pp, p\bar{p}, p\alpha, \alpha\alpha$, etc.), and an antiproton beam was stored for 345
  hours! But nearly all the running was with $pp$ collisions, with center-of-mass energy $\sqrt{s}$
  ranging from 23 GeV to 63 GeV. To reach $\sqrt{s}$ = 63 GeV with a proton beam on a hydrogen target
  would require a beam energy of 2110 GeV, so much higher than the then-record 26-28 GeV of the CERN PS
  and Brookhaven AGS that we were reaching ``\emph{into the realm of the cosmic rays!}''. Not
  only did the machine open a new chapter in physics of relevance to cosmic ray studies, it first
  demonstrated stochastic beam cooling that paved the way for the Sp$\mathrm{\bar{p}}$S and Tevatron proton-antiproton
  colliders, and their discoveries of $W,Z,$ and top-quarks, etc. 
  
  Unfortunately experiments at the ISR did not discover any new \emph{particles}, although both charm and
  bottom quarks were being produced. In stark contrast to the state of preparedness of the LHC detectors,
  when the first collisions occurred in the ISR the only detectors to observe them~\cite{isr1} were a few hastily
  installed scintillation counters and an oscilloscope! Experiment R101~\cite{r101}
  (Rings-Intersection 1, 01) was a child's toy train set with photographic emulsions on each truck.
  Parked alongside the collision region, it measured the polar angle $\theta$ distribution of 
  charged particle production! (The pseudorapidity $\eta$ = -ln tan($\theta$/2) distribution is roughly
  flat.) In Intersection 2, three experiments surveyed $\pi^{\pm}, K^{\pm}$ and $p,\bar{p}$ production at
  small, medium and large polar angles, and a fourth looked for high-$p_T$ muons coming from $W$-decay
  (Perhaps $M(W)$ was only a few GeV/c$^2$!). Five other collision regions were home to a variety of
  experiments, looking for free quarks (which \emph{might} have been abundantly produced), photons and
  electrons, studying particle correlations, etc. While no new particles were discovered, new
  \emph{phenomena} certainly were, from the rising total cross section $\sigma_T$, high $p_T$
  production (including direct photons), high mass diffraction and double pomeron exchange, and so on. 
  
  The ISR experiments that pass my ``relevant-for-cosmic rays'' cuts are
  the forward single- and multi-particle spectrometers. Studies of hadron collisions can be classified as either
  \emph{exclusive} or \emph{inclusive}. In the exclusive case every final state particle is measured; the simplest
  case being elastic scattering: $p+p\rightarrow p+p$. Few-body reactions such as $\pi^- p \rightarrow \pi^0 n$, and 
  low mass diffractive excitation:
  $p+p \rightarrow p+ p\pi^+\pi^-$ can also be fully measured and distributions 
  such as $M(p\pi^+\pi^-)$, $t$-channel variables, and
  decay angular distributions studied.  However even at LHC energies, where most collisions produce a large number of particles, 
  inelastic but exclusive reactions, such as $p+p \rightarrow p + H + p$ can 
  be (rare but) important~\cite{hps}. 
  
  The advent of the ISR, with most interactions having a large particle multiplicity, generated a problem: Events with 10 particles with 40
  variables and only four energy-momentum constraints, and no 36-dimensional graph paper! One popular solution is to just measure one and
  ignore the rest: \emph{inclusive} reactions such as $p+p \rightarrow \pi +$ ``anything($X$)", or eventually, at the Tevatron, 
  $p+\bar{p} \rightarrow W^+,W^-$ (or $Z$, or $t$) etc. + ``anything". The Lorentz-invariant cross section for inclusive pion production
  is $E (d^3\sigma/dp^3) = \sigma_{inv}(s,p_T,x_F) \stackrel{s \rightarrow \infty}\rightarrow \mathit{f}(p_T,x_F)$. The latter limit is
  Feynman's scaling hypothesis~\cite{feynscal}, which preceded the parton model and QCD. At about the same time (1969)
  Benecke, Chou, Yang and Yen proposed~\cite{hlf} the ``hypothesis of limiting fragmentation'', HLF. Consider a high energy particle hitting
  a target particle, and causing the latter to ``fragment'' or create particles, with some distribution in the target frame. The
  HLF hypothesis is that as the projectile energy $E$ gets very large, the target fragments reach limiting distributions in the target
  frame, independent of $E$, and of course the same applies to the projectile fragments in \emph{its} frame. In the target frame the
  projectile fragments then exhibit scaling ($p_z \propto E$) and \emph{vice versa}. The target and
  projectile fragments can be either distinct, and separated by a large gap in rapidity $y$ (double diffraction), or connected by a 
  ``string'' of hadrons (a rapidity plateau). The plateau length has $y_{plateau} \propto \mathrm{ln}\sqrt{s}$, and the multiplicity would rise 
  logarithmically with $\sqrt{s}$.
  A nice (but limited in $\sqrt{s}$ range) demonstration of the HLF was made by experiment R801~\cite{r801} 
  with simple scintillation
  counter hodoscopes, using colliding beams of different momenta. They measured the
  full ($-5 < \eta < 5$) distribution of charged particles in collisions of (in GeV) 15.4+15.4, 26.7+26.7, and 15.4+26.7. The
  produced particles in one direction do not care about the momentum of the opposite-going proton. Now we know that these
  statements, as well as Feynman scaling, are only approximate.

  \begin{figure}[t]
\includegraphics[width=130mm,angle=0]{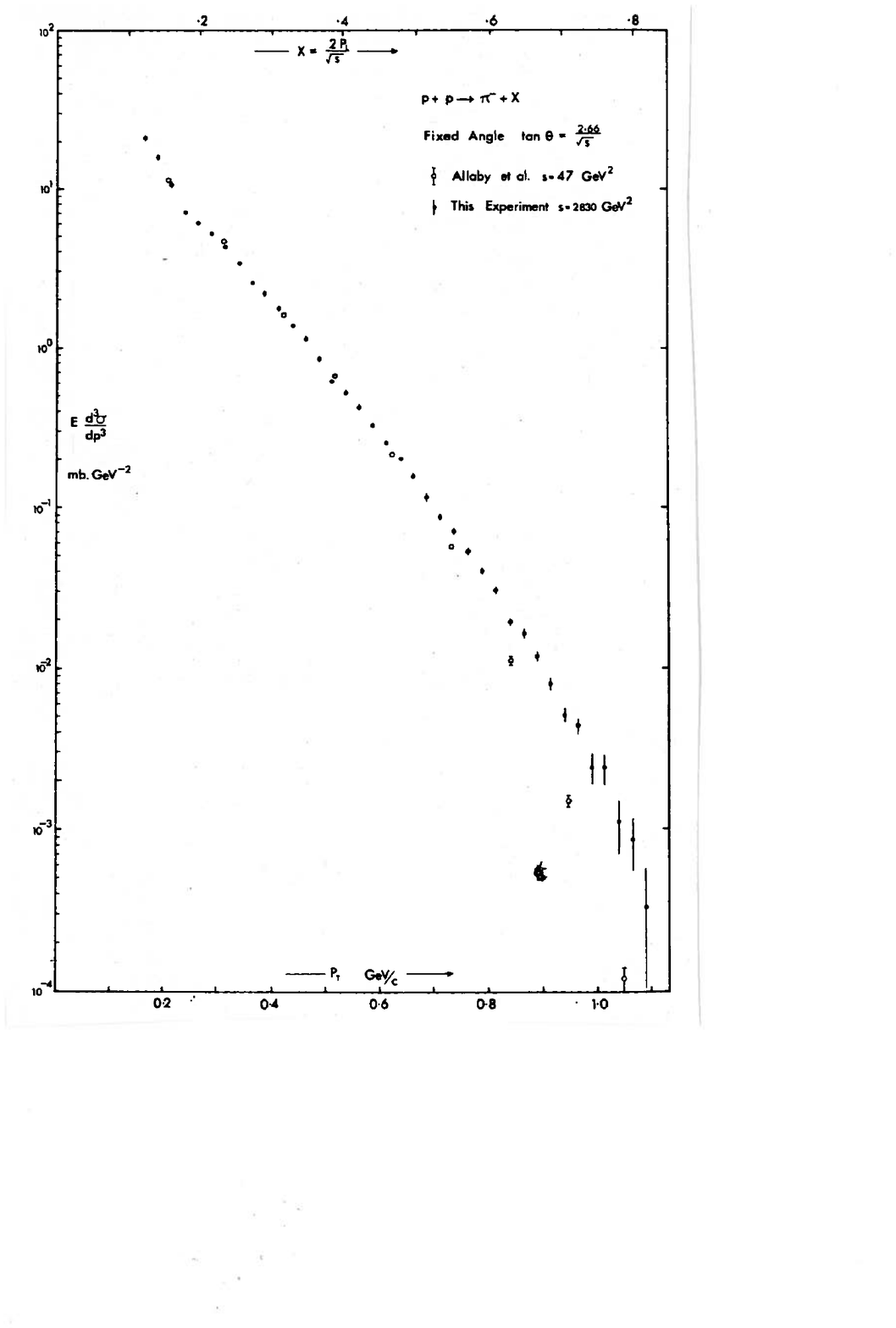}
\vspace{-6.0cm}
\caption{Spectrum of $\pi^-$ at three values of $\sqrt{s}$ at the ISR~\cite{r201},
showing the rise from the PS ($\sqrt{s}$ = 47 GeV). The spectra are at a fixed
angle in the $x_F$ (top axis): $p_T$ (bottom axis) plane.}
\label{saspix}
\end{figure}

  Pre-1971, and in the early days of the ISR, the main focus was on low-$p_T$ production (``\emph{where most of the particles go}'')
  which is conjugate to large distance ($\hbar c = 197$ MeV.fm). This was the justification for the large Split Field Magnet
  (SFM) facility, which had forward dipole fields but, unfortunately, very bad characteristics in the central region (close
  pole pieces and a quadrupole field). By 1976, after the discovery of abundant (relative to most expectations) high-$p_T$ particle
  production, resulting from small distance ($\ll$ 1 fm) parton scattering, the emphasis turned (literally 90$^\circ$)
  to the central region. It has stayed there ever since, measuring parton scattering and in post-ISR colliders testing QCD at distances as small as
  1/1000$^{\mathrm{th}}$ fm, finding $W$ and $Z$ and $t$-quarks, etc. The forward region, $x_F \gtrsim$ 0.05, or $|y-y_{\mathrm{beam}}|<3$,
  became largely neglected. However Experiment R201, the Small Angle Spectrometer~\cite{r201}, SAS, in the first (1971) round of experiments,
  measured the inclusive spectra of $\pi^+,\pi^-,K^+,K^-,p,\bar{p}$ at low-$p_T$, with 0.1 $<x_F<$ 1.0, at several $\sqrt{s}$
  values from 23 GeV - 63 GeV. The 30 m long moveable spectrometer had septum magnets to bend small angle particles away
  from the beam, wire spark chambers, and Cherenkov counters to identify hadrons. I do not have space to show many of the detailed spectra from R201, which can
  be found in the papers~\cite{r201}. A summary follows.
  
\begin{figure*}[t]
\includegraphics[width=135mm,angle=-90]{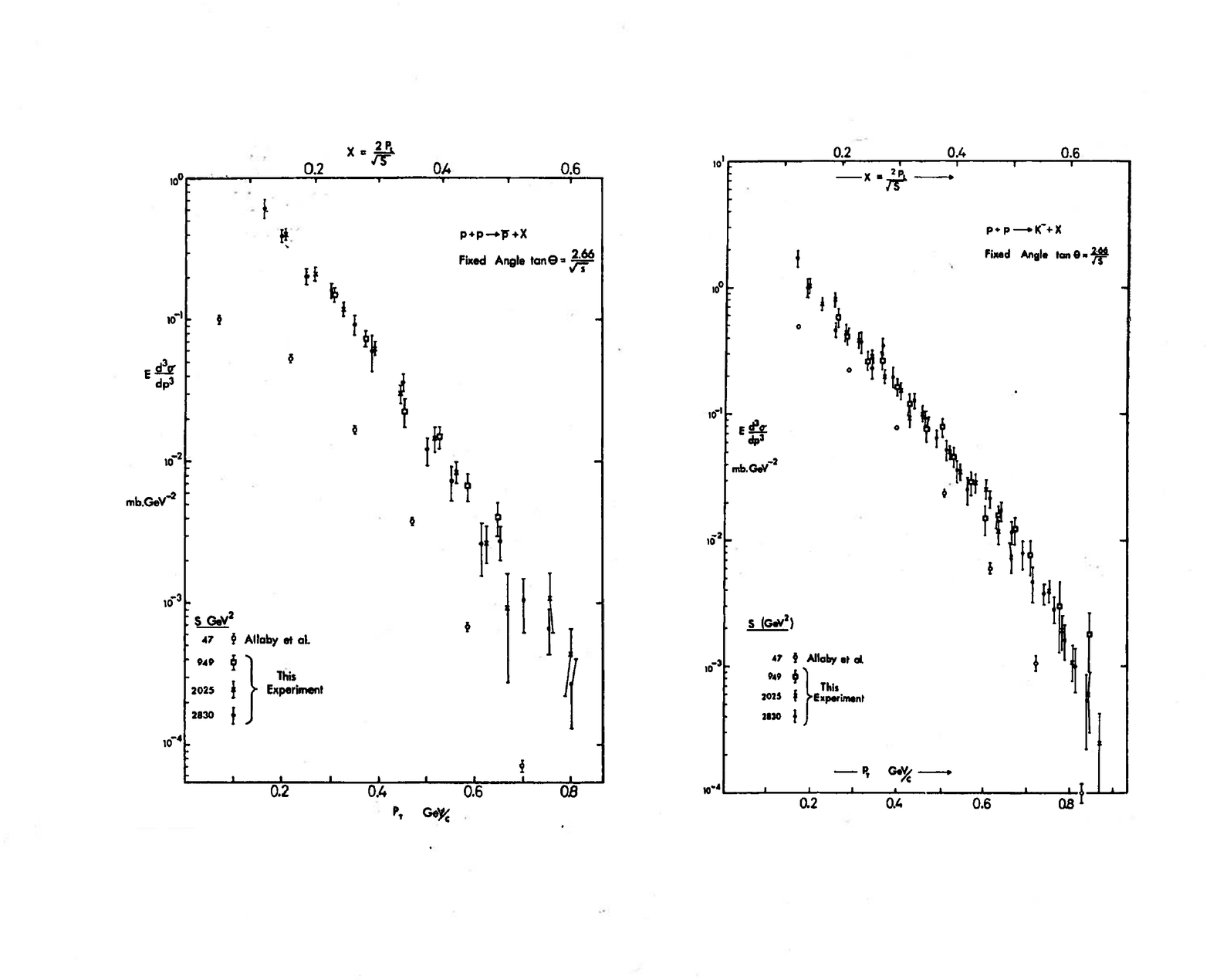}
\vspace{-2.0cm}
\caption{Left: Spectrum of antiprotons~\cite{r201} at three values of $\sqrt{s}$ at the ISR,
showing the rise from the PS ($\sqrt{s}$ = 47 GeV). The spectra are at a fixed
angle in the $x_F$ (top axis): $p_T$ (bottom axis) plane. Right: The same, for
$K^-$.}
\label{saspi}
\end{figure*}

  As the SAS coverage had a $p_T$ range which depends on $x_F$, one could plot the spectra at fixed angle, and vary
  the angle with $\sqrt{s}$, tan $\theta = \frac{2.66 \mathrm{GeV}}{\sqrt{s}}$, so that they are superimposed 
  in the ($p_T,x_F$) plane. The $\pi^-$ spectrum, see Fig.~\ref{saspix}, shows excellent agreement with much lower $\sqrt{s} = 6.8$
  GeV data out to $x_F$ = 0.5, but by $x_F$ = 0.8 (and $p_T$ = 1.0 GeV/c) it is a factor of 
  several higher. 
  The $K^-$ ($\bar{p}$) spectra (Fig.~\ref{saspi}) were a factor $\sim 2 (\sim 13)$ higher than at $\sqrt{s} = 6.8$, but showed
  scaling within the ISR energy range. This was seen as good evidence for (approximate) Feynman scaling below $p_T \sim$ 0.5
  GeV/c, and plotting particle ratios ($\frac{K^-}{\pi^-}, \frac{\bar{p}}{\pi^-}$ \emph{vs} $1/\sqrt{s}$) one could believe
  that some asymptotic limit at $1/\sqrt{s} = 0$ was close. 
  
  In those pre-QCD days Regge theory was applied to the large $x_F$ production of all particles. A proton could turn into a
  leading $\pi^+ (\pi^-)$ by exchanging a $t$-channel $N^* (\Delta^{++})$, or into a $K^+$ by exchanging 
  a  $\Lambda^0$ or $\Sigma^0$. As the
  exchanges are in the $t$-channel, with negative $M^2$, they are virtual, so-called Regge trajectories (sums of states with the
  same quantum numbers), described by a ``spin''~\footnote{Really, complex $t$-dependent angular momentum} $\alpha(t)$.
  E.g. the reaction $p+p \rightarrow \pi^+ + X$ then has the form $\sigma_{inv} = A(t) \left( \frac{M_X^2}{s}  \right)^{1.0 - 2
   \alpha_{N^*}(t)}$. By measuring the $s$- (or $M_X^2$)-dependence at several $t$-values one could map out~\cite{sasregn} the trajectory $\alpha(t)$ and find that
   indeed it fits on a straight line with the real neutron and $N^*$ masses.
   
\begin{figure}[t]
\includegraphics[width=120mm,angle=0]{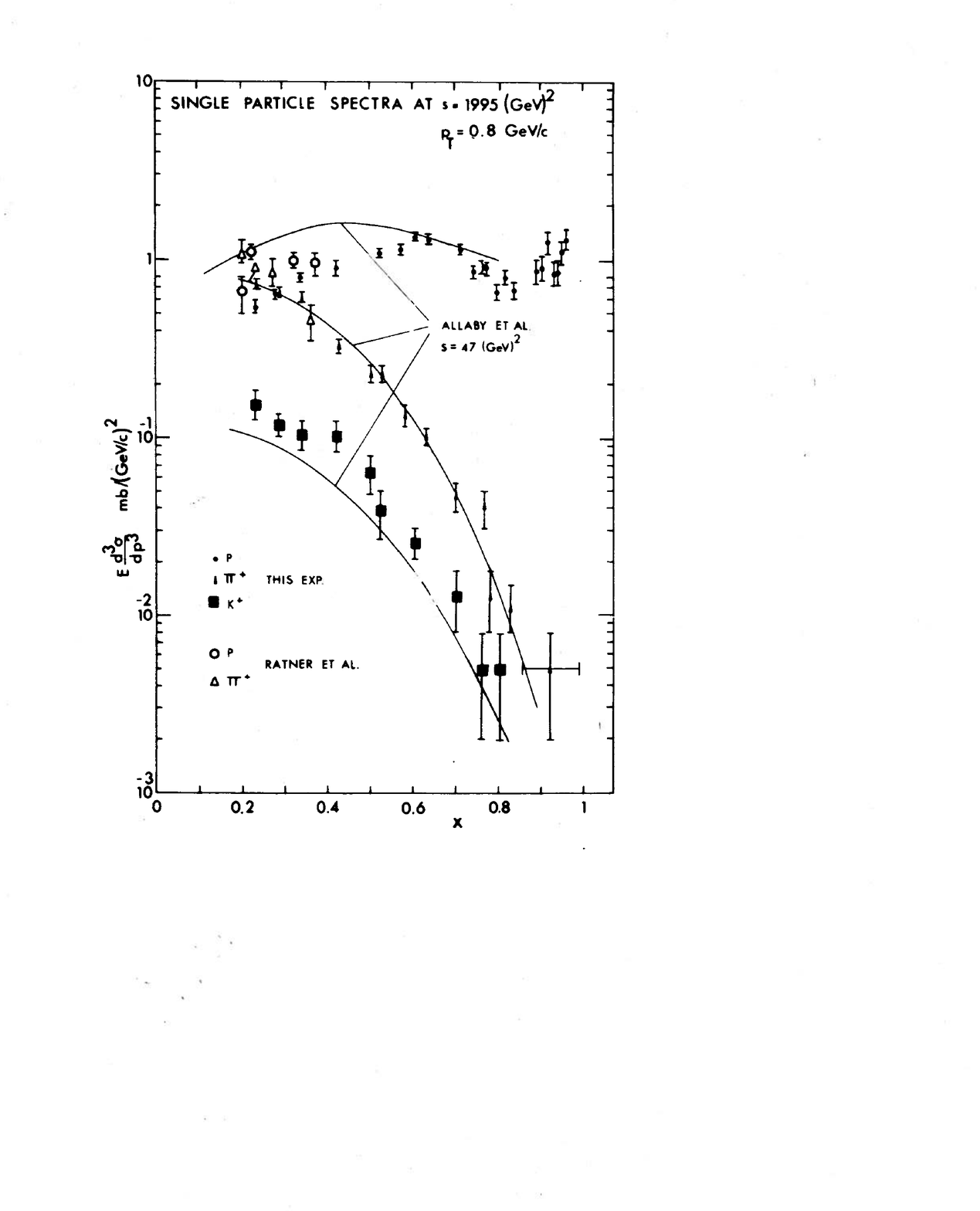}
\vspace{-5.5cm}
\caption{Longitudinal momentum spectra of positive particles at the ISR at $p_T$ = 0.8 GeV/c~\cite{r201}, compared
to $\sqrt{s}$ = 6.8 GeV data (lines). More than 99\% of particles with $x_F >$ 0.8 are protons.}
\label{fig1}
\end{figure}

   The $\pi^+ ,\pi^-$, and $K^+$ spectra showed~\cite{sasplus} similar behavior in $x_F$, with $\pi^+$ measured out to $x_F$ = 0.9 and
   scaling already from $\sqrt{s} = 6.8$ GeV. More interestingly, the proton spectra showed~\cite{r201,saspk} a
   minimum around $x_F = 0.95$, with a high-$x_F$ peak corresponding to diffractive excitation of the opposite proton to a
   state of mass $M_X$, well above the resonance ($N^*$) region, see Fig.~\ref{fig1}. (The data shown in
   Fig.~\ref{fig1} represent the
first observation of high mass diffraction, and were followed by very detailed studies.) From the kinematic relation $M_X^2/s = 1-x_F$, i.e. 
   $M_X = \sqrt{( 1-x_F)}\sqrt{s}$, we see that while the region $x_F > 0.95$ corresponds to $M_X \sim 1.5$ GeV (the $N^*$
   resonance region) at the PS,
   it corresponds to about 14 GeV at the ISR. If the high-$x_F$ peak continues to scale, diffractive excitation should
   extend to about 440 GeV at the Tevatron, and indeed this is approximately true (and it will extend up to about 3000 GeV at
   the LHC-14 TeV). I stress that these are ``soft'' limits; there is no absolute distinction between diffractive
   and non-diffractive events. To avoid model-dependence, both experimenters and theorists should define precisely their criteria, e.g.
   $\sigma(x_F(p)) > 0.95$ or $\sigma(\Delta y > 3.0)$ or similar. However in the context of ``models'', 
   Regge theory described the high-$x_F$ peak as due to the exchange of a ``pomeron'', $\pom$, the same entity that is
   the dominant exchange between two protons in elastic scattering at high energy. Regge theory is based on sound,
   fundamental principles, namely that the scattering amplitudes should obey unitarity, analyticity and crossing symmetry.
   Is it too much to hope that one day it will be unified with QCD into a true ``Theory of Strong Interactions"?
   
   At low energies elastic scattering is dominated by ``reggeon'' exchanges, which are sums of virtual mesons with the
   same quantum numbers. As the exchange is in the $t$-channel, meaning momentum is exchanged but (in one 
   frame) no energy, its angular momentum, $\alpha$, is not real and integer, but complex and varies continuously with
   $t$. The ``Regge trajectory'' $\alpha(t)$ is linear in $t$ and could be mapped out by 
   measuring (in $p+p \rightarrow p + X$) the $M_X^2$-dependence:
   $\sigma(M_X^2,t) \sim (M_X^2)^{1.0-2\alpha(t)}$. In $0.95 < x_F < 0.99$ the measured trajectory is shallow, with slope $\alpha' \sim
   0.2$ and intercept $\alpha(t=0) \sim 1.2$ corresponding to the pomeron, while for $0.5 < x_F < 0.85$ the slope is near 1.0 and
   intercept $\alpha(0) \sim$ 0.5, corresponding to virtual $\rho,\omega$ exchange. The pomeron is to first order
   (and at low $Q^2$) a pair of gluons in a color singlet state~\cite{lown} (but it is always virtual and never isolated like a pion).

     Inclusive neutron spectra at $\theta$ = 0$^\circ$ were measured~\cite{nzero} in a small hadron calorimeter. 
     The identical principle is now used at the LHC
     in the Zero Degree Calorimeters (ZDC) in LHCf, ATLAS and CMS. At fixed $p_T (< 0.5$ GeV/c) the $x_F$-distributions are
     rather flat, but at $\theta = 0^\circ$ there is a distinct bump, described by reggeized pion exchange $p \rightarrow
     n + ``\pi"$. 
     
     Beyond single particle inclusive spectra, forward multiparticle spectrometers such as Expt. 
     R603~\cite{r603} 
     and the Split Field Magnet
     studied fragmentation systematics, and forward strangeness and charm production. These could profit from the ability of the
     ISR to make $p\bar{p}$ collisions. R608~\cite{r608} measured the $x_F$ distributions of both +ve and -ve particles in the fragmentation
     regions of both $p$ and $\bar{p}$, finding that (a) baryon fragmentation is independent of whether the opposite beam is $p$ or
     $\bar{p}$ (Left-Right factorization), and (b) $p \rightarrow h^+ = \bar{p} \rightarrow h^-$ (C-conjugation). Other similar
     equalities were found in $\Lambda$ and $\bar{\Lambda}$ production by fragmenting $p$ and $\bar{p}$ on $p$ or $\bar{p}$
     ``targets''. Protons can fragment diffractively (by $\pom$-exchange) $p \rightarrow \Lambda + K^+$, and somewhat surprisingly the
     $\Lambda$ were found (from their decay distributions) to be polarized. Systematic studies of both $p(uud) \rightarrow \Lambda (uds)$ and
     $p\rightarrow \Delta^{++} (uuu)$ showed that single-quark annihilation is a dominant mechanism at medium $x_F$. 
     
     The proton can also fragment into charmed baryons; $\Lambda_c^+ (cud)(2286)$, was observed~\cite{charmb} through 
     its $\Lambda^+_c \rightarrow \Lambda 3\pi$ decay (2.6\% B.R.). There are predictions~\cite{lambdab} for $p \rightarrow \Lambda_b^0$ at the LHC,
     which in a quark-gluon string model involves exchange of the $\Upsilon$ trajectory, which would be an interesting thing to
     measure. However no existing or foreseen LHC experiments could detect $\Upsilon$s with $x_F
     \gtrsim 0.01$, so we may never
     know!
     
     The Split Field Magnet experiments~\cite{sfm} took advantage of the 
     versatility of the ISR in providing $pp, pd$, and $dd$ collisions, and a
     leading proton from a deuteron could ``tag" the neutron. Thus they could compare, e.g.,
     \begin{itemize}
     \item $pp \rightarrow (p \pi^+\pi^-) + (p \pi^+\pi^-)$

\item $nn \rightarrow (p \pi^-) + (p \pi^-)$ 
\item $pn \rightarrow (p \pi^+\pi^-) + (p \pi^-)$   
 \end{itemize}
     
 showing excellent factorisation; the pomeron does not care about the ($u$ or $d$) quark nature of the fragmenting baryons. This is further
 evidence that the pomeron is gluonic. The cross sections for these specific exclusive reactions rise slowly with $\sqrt{s}$, as does elastic
 scattering.

  %Beyond the single particle inclusive spectra we have 2-particle inclusives,such as $p+p \rightarrow \pi^+ + K^- + X$, 
  %and correlations. Peaks in two particle masses ($D^0 \rightarrow \pi^+ K^-, J/\psi \rightarrow \m^+\mu^-$, etc.) 
  %take one back to single particle spectra!

\section{Fixed target experiments}

While this section takes a step down in $\sqrt{s}$ from the previous one, the
data only came later when high energy beams became available. I select a few
highlights out of a large data set.

The Fermilab Main Injector Particle Production (MIPP) experiment~\cite{mipp} will be
presented at this conference by R.Raja, but it deserves a mention here. It is
designed to do a wide-ranging survey of forward particle production with 120
GeV/c proton beams and secondary $\pi^{\pm},K^{\pm},p^{\pm}$ beams of many
momenta from 5 GeV/c to 85 GeV/c, on a range of target nuclei from H to U. 
The multiparticle spectrometer includes an arsenal of tracking and particle
identification technologies: dE/dx, Time-of-Flight (ToF), and Cherenkov counters. It
would be wonderful if a forward multiparticle spectrometer could be installed at the LHC,
capable of measuring TeV particles!

Since atmospheric cosmic ray interactions are nucleus-nucleus, the ability of
the SPS and RHIC to accelerate Pb nuclei and study Pb-Pb collisions should be
given more prominence in this talk. Searches were made in WA98 (West
Area at CERN)~\cite{wa98} for evidence of unusual events possibly due to a ``disoriented
chiral condensate'', with an extreme charged:neutral particle ratio as had
been reported in cosmic ray interactions (Centauro and anti-Centauro events).
Theoretical ideas suggested that a region of ``pseudo-vacuum'' could be created
with its chiral order parameter misaligned in isospin space from the normal vacuum.
WA98 selected high multiplicity Pb+Pb collisons at 158 GeV/c per nucleon, and counted photons and charged
particles. They found no deviations from the \textsc{venus} Monte Carlo generator.
Other searches were made in $p\bar{p}$ collider experiments (UA1~\cite{ua1cen}, UA5~\cite{ua5cen}, and CDF~\cite{cdfcen}, at higher $\sqrt{s}$ but 
with lower statistics. It remains interesting (and fun) to examine tails of distributions for unexpected
phenomena, but the Centauro effect had been claimed to be not rare ($\sim$ 1\% of events).

The SPS fixed target program included several detectors surveying single- and multi-particle spectra.
NA27~\cite{na27} used the European Hybrid Spectrometer (EHS), which combined a bubble chamber (LEBC) with an electronic
spectrometer. Measurements of $\pi^0$ and $\eta^0$ production in $\pi^- p$ collisions at 360 GeV/c showed a
ratio about 3:1 for $\pi^0 : \eta^0$, and while the $p_T$ spectra have different shapes the $M_T = \sqrt{m^2
+ p_T^2}$ spectra have the same slope. They mapped out the famous ``seagull effect",
which is that $\langle p_T(\pi^0) \rangle$ is minimum at $x_F = 0$, so when plotted over $-1 < x_F < +1$ is
has a seagull shape. 

NA22~\cite{na22} also used the EHS and sent 250 GeV/c $\pi^+$ and $K^+$ beams onto H, Al and Au targets (the latter were
foils in the bubble chamber liquid). Measuring ``Vee"s in the bubble chamber gave the $x_F$ distributions of
$K^0_s, \Lambda^0$ and $\bar{\Lambda^0}$. Strangeness production was found to occur preferentially in central
collisions, and the \textsc{fritiof} Monte Carlo gave reasonable agreement except in the backward region
$x_F \lesssim -0.3$. 

NA61 (SHINE)~\cite{na61} used lower energy $p$-beams, around 30 GeV/c, specifically to study hadron production for
cosmic ray and neutrino experiments (e.g. the T2K neutrino beam in Japan, and MINOS at Fermilab). Particle
identification was done with dE/dx in a time projection chamber and ToF measurements. Flight paths in
collider experiments are very limited in path length, but in SHINE 13 m could be used to extend the
range, and produce $\pi^{\pm}$ and $K^{\pm}$ spectra with good statistics up to about 6 GeV/c ($x_F \sim$
0.2). NA49~\cite{na49} used the same detectors with Pb+Pb collisions with 158 GeV/A, to look 
for unusual events with very
large or very small $\langle p_T \rangle$. They calculated the $\langle p_T \rangle$ distribution for events of a fixed
multiplicity, and compared it with a ``mixed event'' distribution, made from random tracks from a number of events with the
same multiplicity. Again, nothing unusual was seen at the level $\sim 10^{-3}$.

\section{CERN Sp$\mathsf{\bar{p}}$S Collider}
The Sp$\mathrm{\bar{p}}$S collider gave us the factor $\times$10 step up in
 $\sqrt{s}$ that enabled the $W$- and $Z$-bosons to be discovered, and
 high-$E_T$ jets from quark and gluon scattering to be abundantly produced 
 (the jets were co-discovered at the ISR, but they were much more prominent at the 
 Sp$\mathrm{\bar{p}}$S). The big central experiments, UA1 and UA2, did not measure forward particles.
 
  A remarkable general survey experiment was the 6 m long
 streamer chamber of UA5~\cite{ua5}, which extended to $\theta_{min} = 0.6^\circ$. When hits 
 were detected in scintillation counter or Pb-glass
 trigger hodoscopes, a 500 kV pulse was applied to a gas for just 10 ns, causing discharges along the ionization
 tracks that were photographed in stereo, for later scanning. The detector was first commissioned at
 the ISR, and at the Sp$\mathrm{\bar{p}}$S data could be taken from $\sqrt{s}$ = 200 GeV
 up to 900 GeV in a special ``ramping run'' (the SPS magnets could go to 900 GeV only in short bursts without
 overheating). The very detailed spatial information allowed measurements of kaon production using $K^0_s \rightarrow \pi^+\pi^-$ and $K^+
 \rightarrow \pi^+\pi^+\pi^-$ decays, and were compared to model predictions (\textsc{dpm,fritiof,pythia}),
 finding reasonable agreement. The quantities $\langle p_T \rangle, \langle n_K \rangle, \langle K/\pi \rangle$ all
 rise with $\sqrt{s}$, with $\langle K/\pi \rangle$ = 0.11 (in $|y| <$ 3.5) at $\sqrt{s}$ = 900 GeV. UA5 was also able
 to measure photons using conversions in the vacuum pipe or in a Pb-glass plate inserted for that
 purpose. Most photons come from $\pi^0$-decay, with some from $\eta$-decay. There was no sign of events with
 an unusual ratio $\gamma/\pi^{\pm}$ as had been suggested from cosmic ray data (Centauros). A study of
 KNO (Koba-Nielsen-Olesen) scaling, in which $\langle n_{ch} \rangle \times P_n$ ($P_n$ being the probability of $n$
 charged particles) depends only on $z =
 n/\langle n \rangle$  was found to be reasonable, but not exact, over $\sqrt{s}$ = 200 GeV - 900 GeV.

\section{Fermilab Tevatron p$\mathrm{\bar{p}}$ Collider}

The Sp$\mathrm{\bar{p}}$S collider gave way to the Tevatron, with $\sqrt{s}$ = 1800 GeV and later 1960 GeV.
Highlights of the two main central experiments were the discovery of the top quark, frontier $b$-physics and
amazingly good agreement between high-$E_T$ jet spectra (up to about 800 GeV) and QCD Monte Carlos (after some
tuning). The forward region $0.05 < x_F < 0.85$ was left uncovered, but small trackers in Roman pots measured~\cite{cdfpk}
diffractively scattered forward protons, the $x_F \gtrsim$ 0.95 peak. The total cross section measurements (at 1800 GeV)
span the range $\sigma_T =$ 72-80 mb, unfortunately with a 2$\sigma$ discrepancy between experiments, and elastic
scattering is $\sigma_{el} \sim$ 16-20 mb. A study of interest for cosmic ray physics was a search for Centauros in
CDF, using an open (``zero-bias'') trigger. One looked~\cite{cdfcen} in high multiplicity and/or high $\langle E_T \rangle$ events
for extreme hadronic:electromagnetic ratios in the calorimeters, deriving limits on a distinct class $\lesssim 10^{-4} \times
\sigma_{inel}$. A more focused search was carried out by the T864 (T = Test) MiniMax experiment~\cite{minimax}. 
They measured the $\pi^0:\pi^{\pm}$ ratio out to $\eta \sim$ 4.1 as 
a search for a ``disoriented
chiral condensate''. Interesting ideas (not implemented) were to cover even higher $\eta$ by displacing 
the bunch collision region in $z$, and to use a Tevatron dipole 
as a spectrometer magnet for forward
 $\pi^+$ and $\pi^-$.

\section{RHIC at Brookhaven National Laboratory}

Intermediate in energy between the Sp$\mathrm{\bar{p}}$S and Tevatron colliders
is the Relativistic Heavy Ion Collider, RHIC, at Brookhaven. The main focus is
on heavy ion collisions and searches for phase transitions (quark-gluon
plasma?), and while this is obviously of great relevance for cosmic ray showers
I am not expert and I refer to the talk of Balantekin at this symposium. I will
just comment on some results from $pp$ running at $\sqrt{s} =$ 62.4 GeV (as at
the ISR) and 200 GeV, where
the BRAHMS experiment~\cite{brahms} included forward spectrometers with ToF and a RICH detector to
identify high momentum particles. From $y$ = 0 to 3.8, and
$p_T$ from 0.2 to 4 GeV/c, the
spectra of $\pi^{\pm}, K^{\pm}, p$ and $\bar{p}$ have been measured and compared
with \textsc{pythia}. They find that proton fragmentation in \textsc{pythia}
needs improving, presumably in its treatment of diffraction (and of relevance
to cosmic ray showering).

\section{Large Hadron Collider, LHC}

This section will be brief, as we have talks from all the LHC experiments. The
data are now coming, with $\sqrt{s} = 7$ TeV, 3.6 $\times$ the Tevatron, albeit
with much lower luminosity. Note in particular that the famous ``knee'' in the
cosmic ray spectrum is just in-between the Tevatron and LHC energies, and there
have been suggestions~\cite{white,knee} that it is caused by a change in the \emph{nature} of the
collisions rather than (or in addition to) a change in the flux. Such a change
would probably have to be so dramatic that it could be seen already, so it is
unlikely. While the beautiful and impressive LHC detectors cover nearly all of
4$\pi$ solid angle, they still miss the $x_F >$ 0.05 region ($p_L >$ 175 GeV/c with 3.5
TeV beams, i.e. $<$20 mrad for $p_T$ = 350 MeV/c). The exceptions are TOTEM, with very
forward proton detectors in Roman pots, and the $0^\circ$ calorimeters of LHCf and
ZDC in ATLAS and CMS. There is a proposal to add scintillation counters along
the beams pipes around CMS, the Forward Shower Counters, FSC~\cite{fsc}. These cannot
directly measure medium $x_F$ particles, but these hit the beam pipes and
surounding material and make showers which can be detected. One can
compare the patterns of showers with that expected by event generators
that include diffraction, such as \textsc{dpmjet}, and perhaps tune them.
The FSC would increase the detector coverage close enough to $4\pi$ that, if the
luminosity is known, $\sigma_{inel}$ can be measured, and they have many applications
in diffraction. Hopefully these will be approved and installed in early 2011.

Further forward, both along the beam lines and in time, there is a proposal~\cite{hps} to
add very high precision (1 $\mathrm{\mu}$rad tracking, 10 ps timing) proton spectrometers 
to both ATLAS and CMS. Exclusive Higgs boson production, $p+p \rightarrow p+H+p$, and $W$-pair
production $p+p \rightarrow p+ W^+W^- +p$ should be detectable, and the properties of the $H$ (e.g.)
studied in a unique way. It has been predicted~\cite{white} that $\pom + \pom \rightarrow W^+W^-$
might be much more common than in the Standard Model. 
If true, the implications for cosmic ray showers above $10^{17}$ eV would be dramatic.

Let me close with a question: ``Could one make a forward spectrometer for the LHC capable of
measuring and identifying charged hadrons with $0.05 < x_F < 0.90$?'' (Neutrons and $K^0_L$ are detected in the
ZDC, but not distinguished.) The lack of long straight sections around the
collision regions makes it difficult. Nothing has been worked out, as far as I know, but possibly
one could extract very small angle particles, after the BMX dipoles, using crystal channeling.
There is a straight section, not cryogenic, $\sim$ 60 m long, in which silicon tracking and perhaps
particle identification with transition radiation could be installed. This would be a ``high cross section''
experiment, perhaps with short runs at low luminosity, and so it might even be able to use the idea of
displacing the collision region in $z$ to change the acceptance. This is just ``food for thought''; if it looks
feasible the main difficulty might be the near-100\% focus of experimenters on the central region ($|\eta| <
4$ at the LHC).

\bigskip % extra skip inserted
\begin{acknowledgments}
I thank the DOE for support, and my colleagues especially from the SAS at the ISR with whom I measured high-$x_F$
particle spectra nearly 40 years ago!
\end{acknowledgments}

\bigskip % extra skip inserted

\end{document}